**Datasets of the solar quiet (Sq) and solar disturbed (SD) variations of the geomagnetic field from the mid latitudinal Magnetic Observatory of Coimbra (Portugal) obtained by different methods**

**Anna Morozova[1,2], Rania Rebbah[1], Paulo Ribeiro[2]**

1 University of Coimbra, CITEUC, Department of Physics, Coimbra, Portugal
2 University of Coimbra, CITEUC, OGAUC, Coimbra, Portugal

### Corresponding author
Anna Morozova (anna.morozova@uc.pt, annamorozovauc@gmail.com)

### Abstract
Here we present datasets of daily variation obtained from the geomagnetic field raw observations recorded at the Coimbra Magnetic Observatory (COI, Portugal) between 01.01.2007 and 31.12.2017, covering almost the entire solar cycle 24. Two methods were used to extract daily variability from the raw geomagnetic hourly data. The first method uses the so-called "geomagnetically quiet days" to calculate S-type variations as daily means resulting in the data sub-set named "IQD Sq and SD". The second method uses the principal component analysis (PCA) to decompose the original series into main variability modes. The first three modes produced by PCA and explaining up to 98% of the variability of the raw data are in the data sub-set named "PCA modes". Both methods allow to extract regular geomagnetic field variations related to daily variations (S-type variations) in the ionospheric dynamo region and some magnetospheric currents (e.g., field-aligned currents).

The COI location in middle latitudes near the mean latitude of the ionospheric Sq current vortex's focus allows studying its seasonal and decadal variability using the S-type regular variations of the geomagnetic field measured near the ground. The S-type variations for the X and Y components of the geomagnetic field obtained at the COI observatory can also be re-scaled and used to analyze geomagnetic field variations obtained at other European geomagnetic observatories at close latitudes. The S-type variations for the Z component of the geomagnetic field obtained at the COI observatory can be compared to similar variations observed at more continental regions to study the so-called "coastal effect" in the geomagnetic field variations.

The dataset described in this paper is analyzed in a companion paper "Comparison of the solar variations of the geomagnetic field at the Coimbra Magnetic Observatory (COI) obtained by different methods: effect of the solar and geomagnetic activity" by A. Morozova and R. Rebbah submitted to Adv. Space Res.

**Specifications Table**

| | |
|---|---|
| **Subject** | Geophysics |
| **Specific subject area** | Geomagnetic field variations; regular variations of the geomagnetic field; solar quiet variations (Sq) |
| **Type of data** | ASCII TAB-delimited data files<br>Figures |
| **How data were acquired** | The hourly dataset of the orthogonal components of the geomagnetic field (X, Y and Z) was acquired at the Magnetic Observatory of Coimbra (COI) from 01.01.2007 to 12.31.2017 according to the standard procedures of the International Association of Geomagnetism and Aeronomy (IAGA). The data are in nT.<br><br>Instruments used for acquiring the raw data: (1) absolute observations of Declination and Inclination made with a DI-flux magnetometer (based on a fluxgate Mag-01H sensor mounted on a steel-free universal theodolite YOM MG2KP, Bartington), and an Overhauser GSM-90F1 scalar magnetometer (GEM systems) for the Total field (F); (2) a triaxial fluxgate variometer FGE (Ver. J, DMI) was used for the continuous recordings of H, D and Z components. |
| **Data format** | Analyzed<br>Raw data are available from the World Data Center |
| **Parameters for data collection** | The data were produced from the raw observations data using two different approaches:<br>1. using 5 international quiet days (IQD) provided by the GeoForschungsZentrum (GFZ) Potsdam<br>2. using the principal component decomposition (PCA) |
| **Description of data collection** | The raw data of the geomagnetic field variations (1h cadence) from 01.01.2007 to 31.12.2017 were divided into monthly blocks. Each monthly block was treated separately. Two types of Sq variations were calculated:<br>1. $Sq_{IQD}$ and $SD_{IQD}$ calculated using the IQD for each of the 12*11 months;<br>2. PCA modes (1st, 2nd and 3rd modes) |
| **Data source location** | Institution: University of Coimbra, Geophysical and Astronomical Observatory, Coimbra Magnetic Observatory<br>City: Coimbra<br>Country: Portugal<br>Latitude and longitude for collected data: 40º 13' N, 8º 25.3' W, 99 m a.s.l. |
| | Primary data sources: The raw geomagnetic field data can be downloaded from the World Data Centre for Geomagnetism, Geomagnetism Data Portal (http://www.wdc.bgs.ac.uk/dataportal/), station name: "Coimbra", IAGA code: "COI" |
| **Data accessibility** | Repository name: Mendeley Data<br>Morozova, Anna; Rebbah, Rania; Ribeiro, Paulo (2021), "Datasets of the solar quiet (Sq) and solar disturbed (SD) variations of the geomagnetic field at a midlatitudinal station in Europe obtained by different methods", Mendeley Data, V1<br>Data identification number: doi: 10.17632/jcmdrm5f5x.1<br>Direct URL to data: http://dx.doi.org/10.17632/jcmdrm5f5x.1 |
| **Related research article** | **Companion paper: "**Comparison of the solar variations of the geomagnetic field at the Coimbra Magnetic Observatory (COI) obtained by different methods: effect of the solar and geomagnetic activity" by A. Morozova and R. Rebbah, co-submition to Adv. Space Res. |

## Value of the Data

- Here we provide a secondary data representing regular (daily) variations of the geomagnetic field measured at the ground level for the mid-latitudinal European region and a station located near the latitude of the focus of the ionospheric Sq current vortex.
- Since these Sq and SD data reflect conditions in the ionosphere and magnetosphere both geomagnetic and ionospheric scientific communities can benefit from these datasets.
- These data can be used to study regular daily variations of the geomagnetic field for the mid-latitudinal European region, particularly latitudes near the focus of the ionospheric Sq current vortex.
- Since Sq variation is related to the current ionospheric systems, this data set can be used to study the ionospheric Sq current vortex variability.
- These data can be used directly (especially the X and Y components) in studies of the geomagnetic field disturbances at middle latitudes in the European sector: the Sq variation must be removed from the geomagnetic observations data prior to the analysis of any disturbances.
- Due to the proximity of the Coimbra Magnetic Observatory to the ocean coast, a part of these datasets (Z component of the geomagnetic field) can be used to study the coastal effect on the geomagnetic field variations measured at the ground level.

## Data Description

The dataset consists of two main parts: one of the two subsets is the data on the S-type variations of the geomagnetic field obtained using quiet days (see section "Experimental Design, Materials and Methods/*Quiet days Sq and SD*“ for more details); another subset is the modes of the geomagnetic field variation obtaining with the principal component analysis (see section “Experimental Design, Materials and Methods/Principal component analysis “for more details). These two subsets can be found in the folders “IQD Sq and SD\” and “PCA modes\”, respectively. Each of these folders contains subfolders “X\”, “Y\” and “Z\” containing the data for the corresponding components of the geomagnetic field. Each of the components’ sub-folders, in turn, contains 12 sub-sub-folders (e.g. “m01 – January\”) with data related to each calendar months. This structure is shown in Fig. 1.

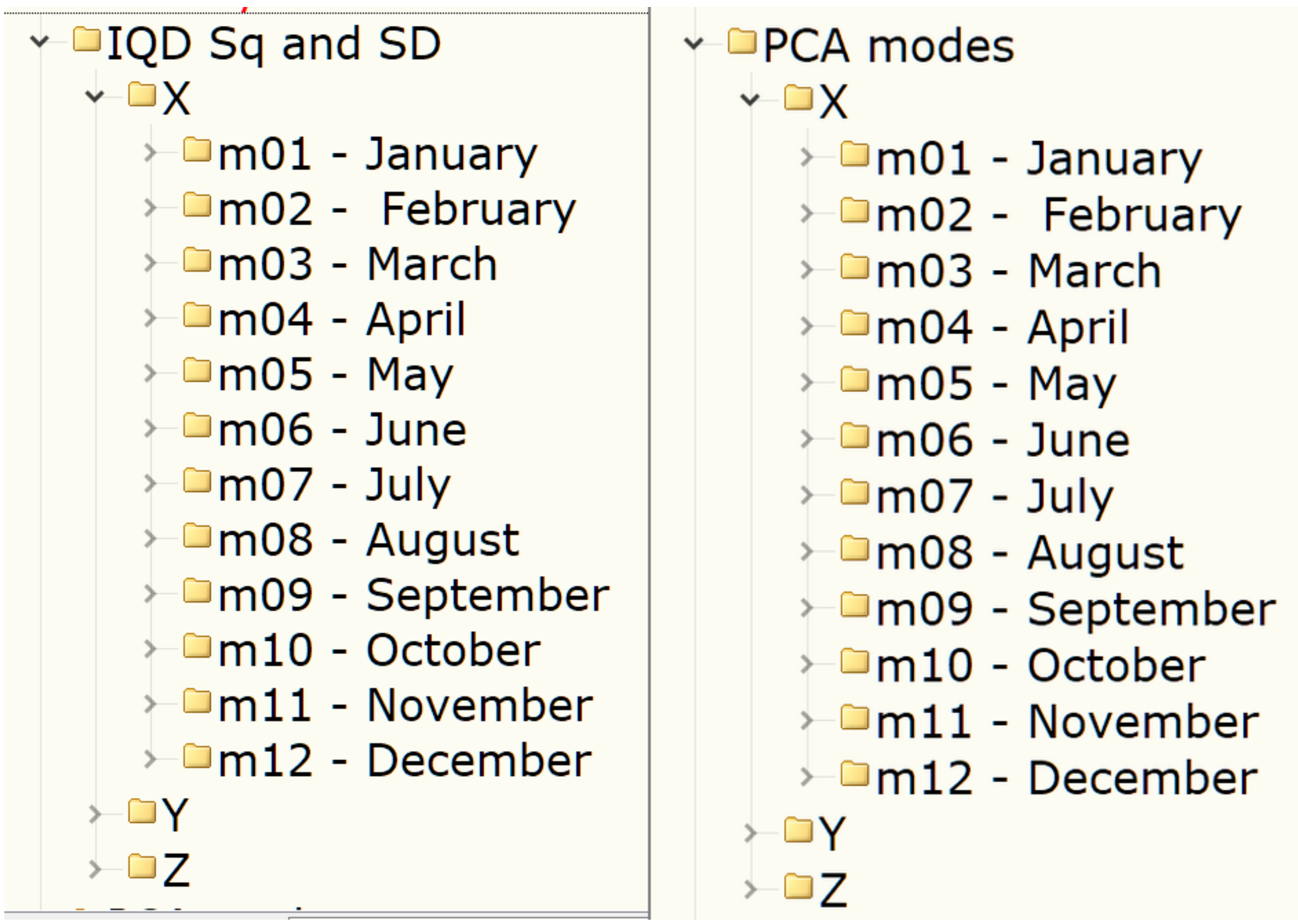

**Figure 1.** Organization of the data in folders: left – folders structure in the folder “IQD Sq and SD\”; right – folders structure in the folder “PCA modes\”.

Each of the monthly sub-sub-folders contains the sub-sub-sub-folders “data files\” and “plots\”. They contain data files (ASCII TAB-delimited *.dat files) and figures (*.png files), respectively. The internal structure and the files placed in the “data files\” and “plots\” folders depend on the subset (either “IQD Sq and SD” or “PCA modes”), as is shown in Fig. 2.

In the case of the “IQD Sq and SD” subset (Fig. 2, top), the “data files\” folders have no sub-folders and contain data files named like “COI_*_Sq.SD.S_m##.dat” where “*” is for “X”, “Y” or “Z” and “m##” is for a month (please also see Table 1 for a list of nomination and abbreviations used to name the files). These data files contain Sq, SD and S variations for a certain month (“m##”) and different years (y.20##) or for all years “y.all” (see Fig. 3). S-type variations are in nT and calculated as described in section “Experimental Design, Materials and Methods/*Quiet days Sq and SD*”.

**Table 1.** List of nomination and abbreviations used to name the data and figure files

| Nomination/abbreviation | Description |
|---|---|
| COI | IAGA code of the observatory |
| X, Y or Z | Component of the geomagnetic field |
| m01, m02, ..., m12 | Data are related to a certain month of a year |
| y.2007, ..., y.2017 | Data are related to a certain year |
| year_2007,..., year_2017 | |
| y.all | Data are related to all analyzed years (2007-2017) |
| y_all | |
| Sq, SD or S | Type of the S variation |
| PC1, PC2 or PC3 | Principal component representing certain daily variations of the field |
| EOF1, EOF2 or EOF3 | The amplitude of a corresponding PC for a certain day of a month |
| mode 1, mode 2 or mode 3 | Reconstructed mode |
| mode1 or mode2 | |
| VF1, VF2 or VF3 | Variance fraction associated with a certain mode |
| SCF1, SCF2 or SCF3 | |

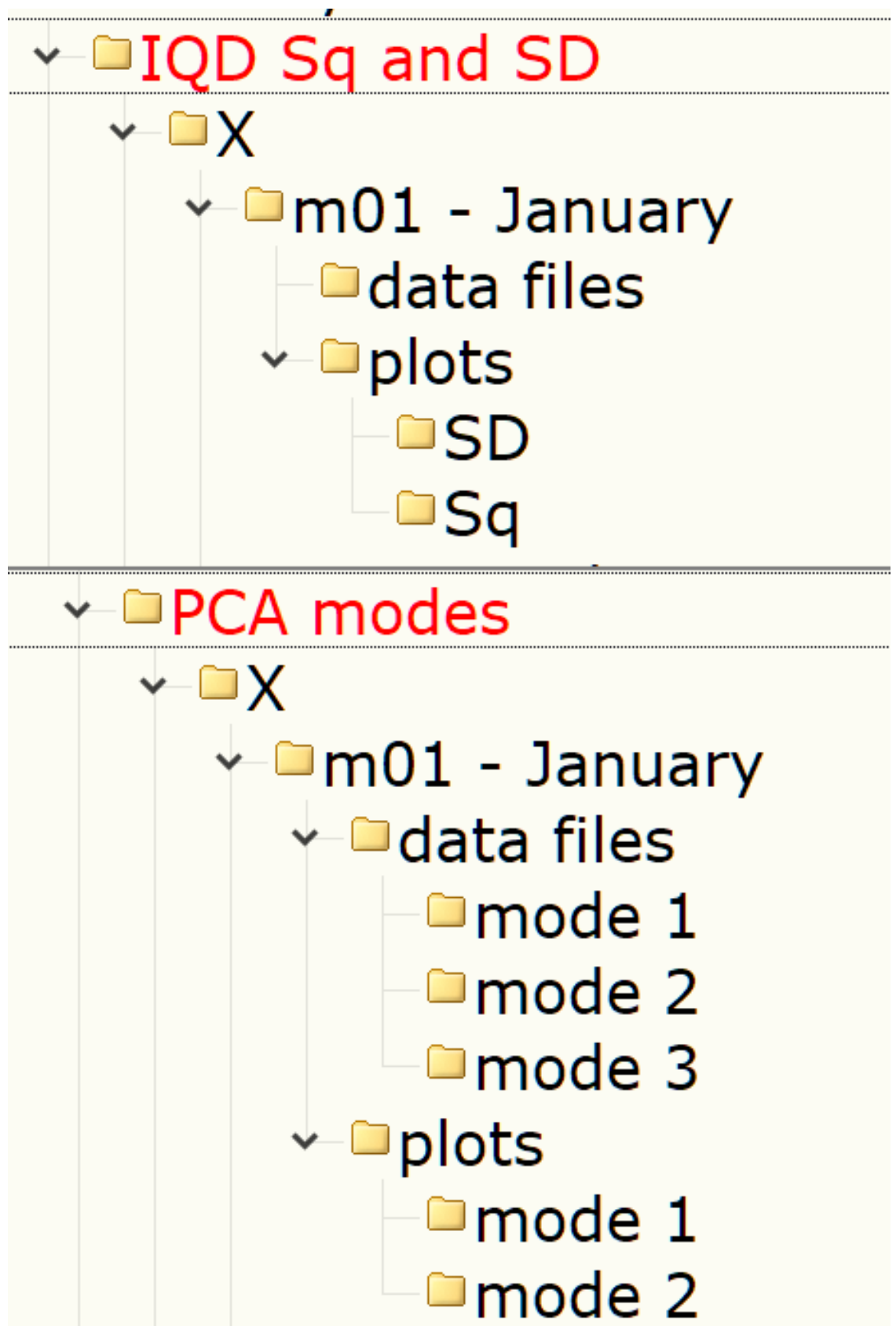

**Figure 2.** Examples of the data's organization in monthly folders: top – sub-folders structure in the folder "IQD Sq and SD\X\m01 - January"; bottom – sub-folders structure in the folder "PCA modes\X\m01 - January".

The "plots\" folders contain two sub-folders named "Sq" and "SD" that have figures with corresponding plots of the Sq and SD variations, respectively. The files are designated as "COI_*_S*_m##_year_20##.png" where "*" is for "X", "Y" or "Z", "S*" if for a S-type, "m##" is for a particular month and "year_20##" is for a specific year.

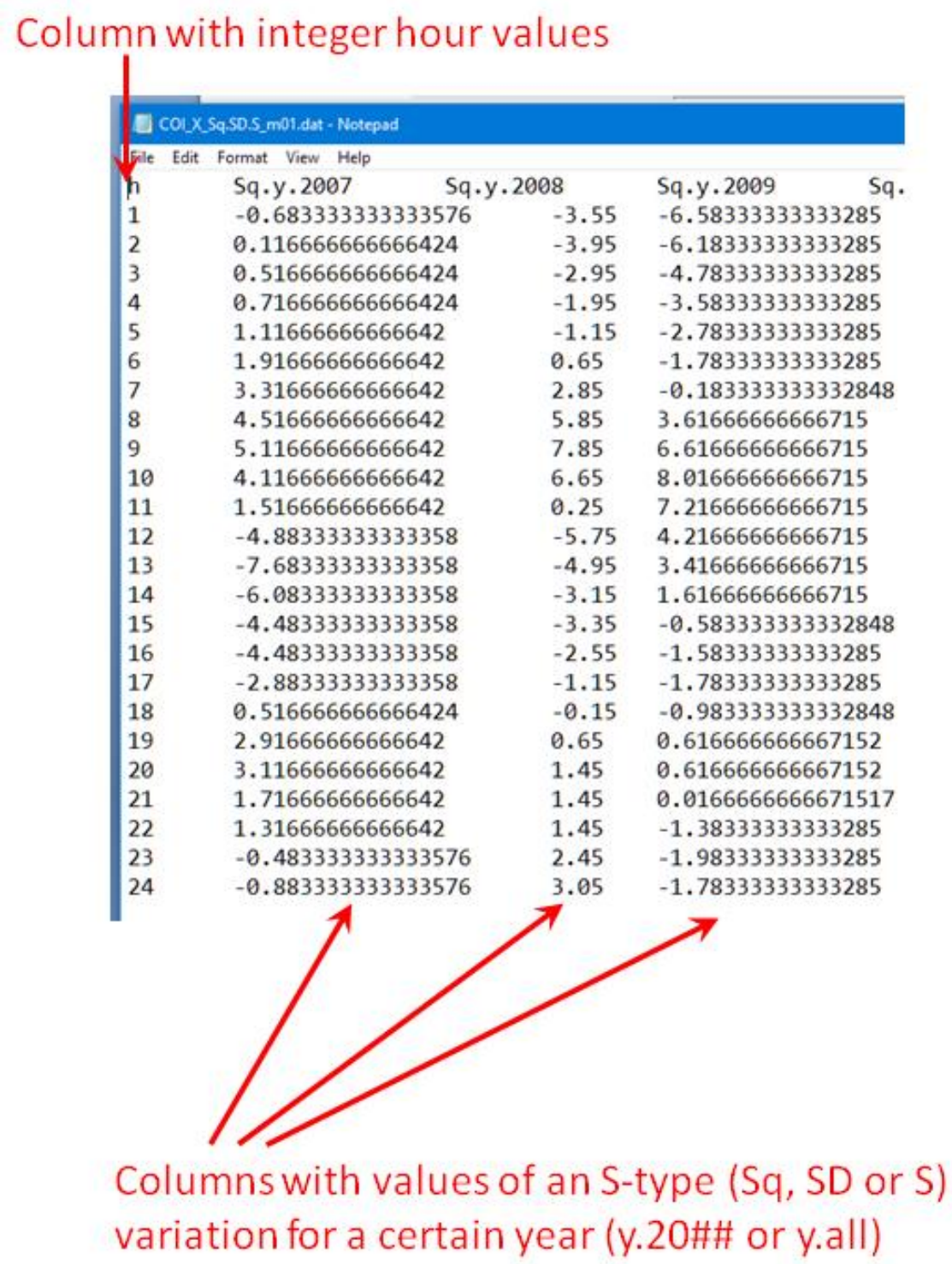

**Figure 3**. Internal structure of the "COI_*_Sq.SD.S_m##.dat" files.

In the case of the "PCA modes" subset (Fig. 2, bottom) each of the "data files\" folders contain three sub-folders "mode 1\", "mode 2\" and "mode 3\". Each of the "mode" sub-folders have data files related to a respective PCA mode, see Fig. 4. These files are names as follow: "COI_*_EOF#_m##.dat" (EOF series for individual years) and "COI_*_EOF#_y_all_m##.dat" (EOF1, EOF2 or EOF3 series for all analyzed years); "COI_*_PC#_m##.dat" (PC1, PC2 or PC3 series for individual years and for the all 11 years together); "COI_*_mode#_m##.dat" and "COI_*_mode#_m##_y_all.dat" (reconstructed mode 1 or mode 2 for individual years and all analyzed years, respectively), "COI_*_VF#_m##.dat" (variance fraction related to this mode for individual years and for the all 11 years together) where "*" is for "X", "Y" or "Z", "m##" is for a month, and "VF#" is for a variance fraction for a mode# (please also see Table 1). Please note that reconstructed modes are calculated only for the 1$^{st}$ and 2$^{nd}$ modes. All series are either in nT (EOF), in arbitrary units (PC, mode) or non-dimensional (VF) and calculated as described in the section "Experimental Design, Materials and Methods/Principal component analysis" for more details. The internal structure of the "*PC*.dat", "*EOF*.dat", "*mode*.dat" and "*VF*.dat" files are shown in Figs. 5-8. Also, the "data files\" folders contain two files with daily mean values of a selected component for a particular month: "COI_*_meanField_m##.dat" with data for individual years and "COI_*_meanField_m##_y_all.dat" with data for all analyzed month. Their internal structure is shown in Fig. 9.

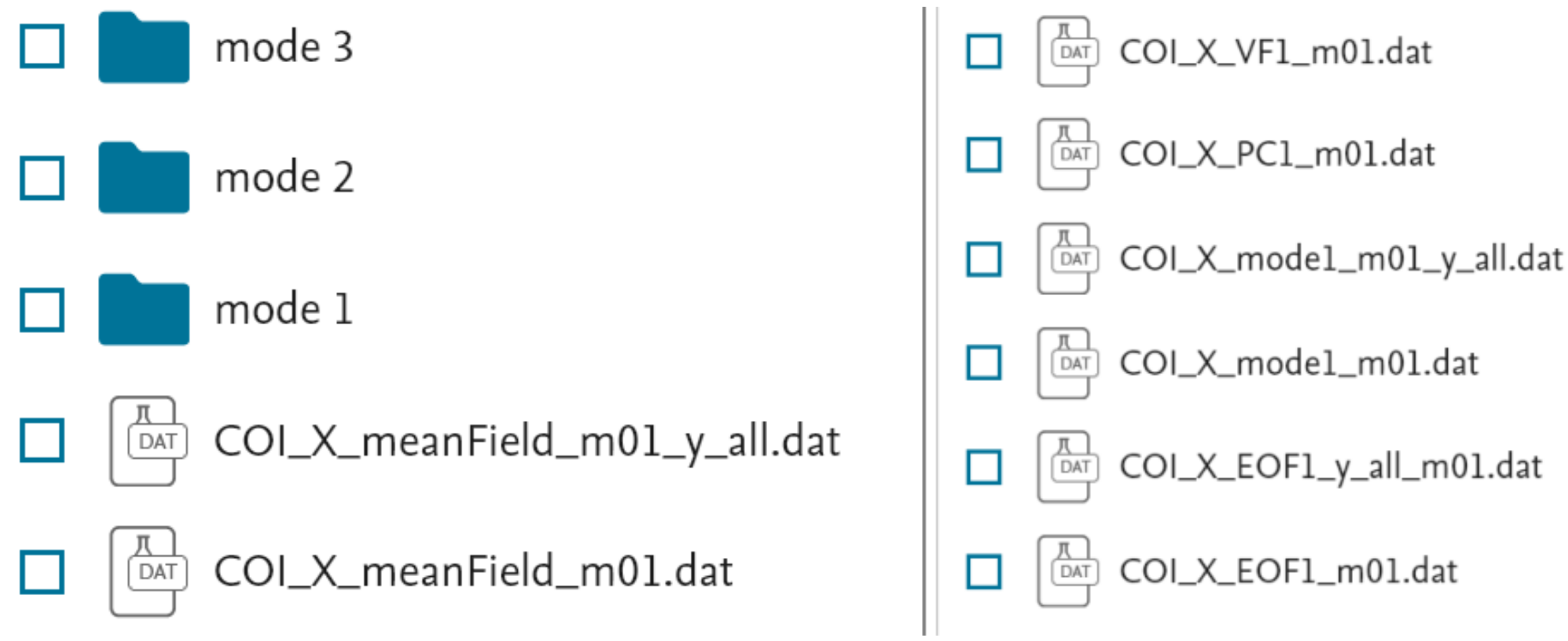

**Figure 4.** Examples of the data's organization in monthly data folders: left – sub-folders structure in the folder “PCA modes\X\m01 – January\data files”; right – files in the folder “PCA modes\X\m01 – January\data files\mode 1”.

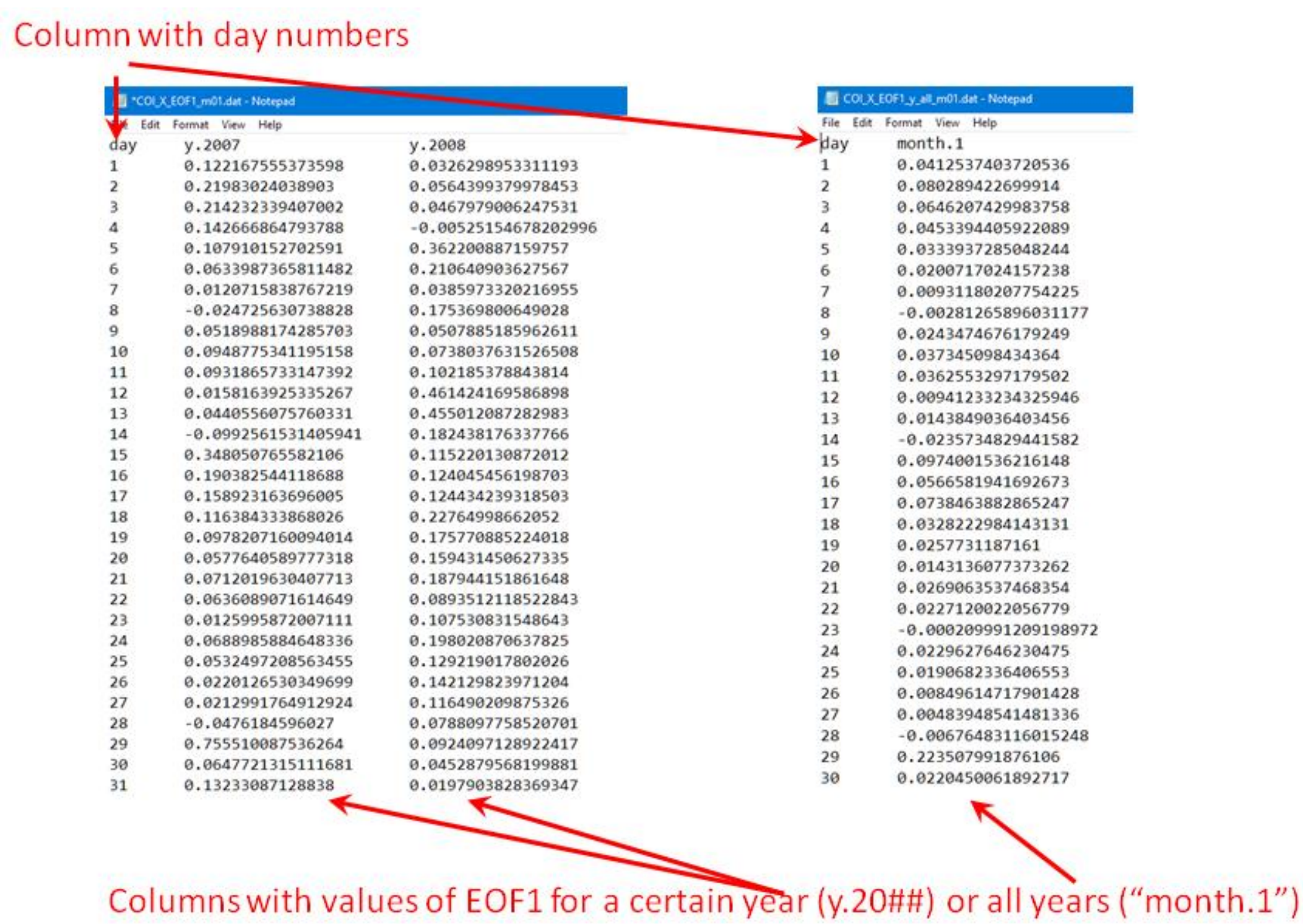

**Figure 5.** The internal structure of the “COI_*_EOF#_m##.dat” (left) and “COI_*_EOF#_y_all_m##.dat” (right) files.

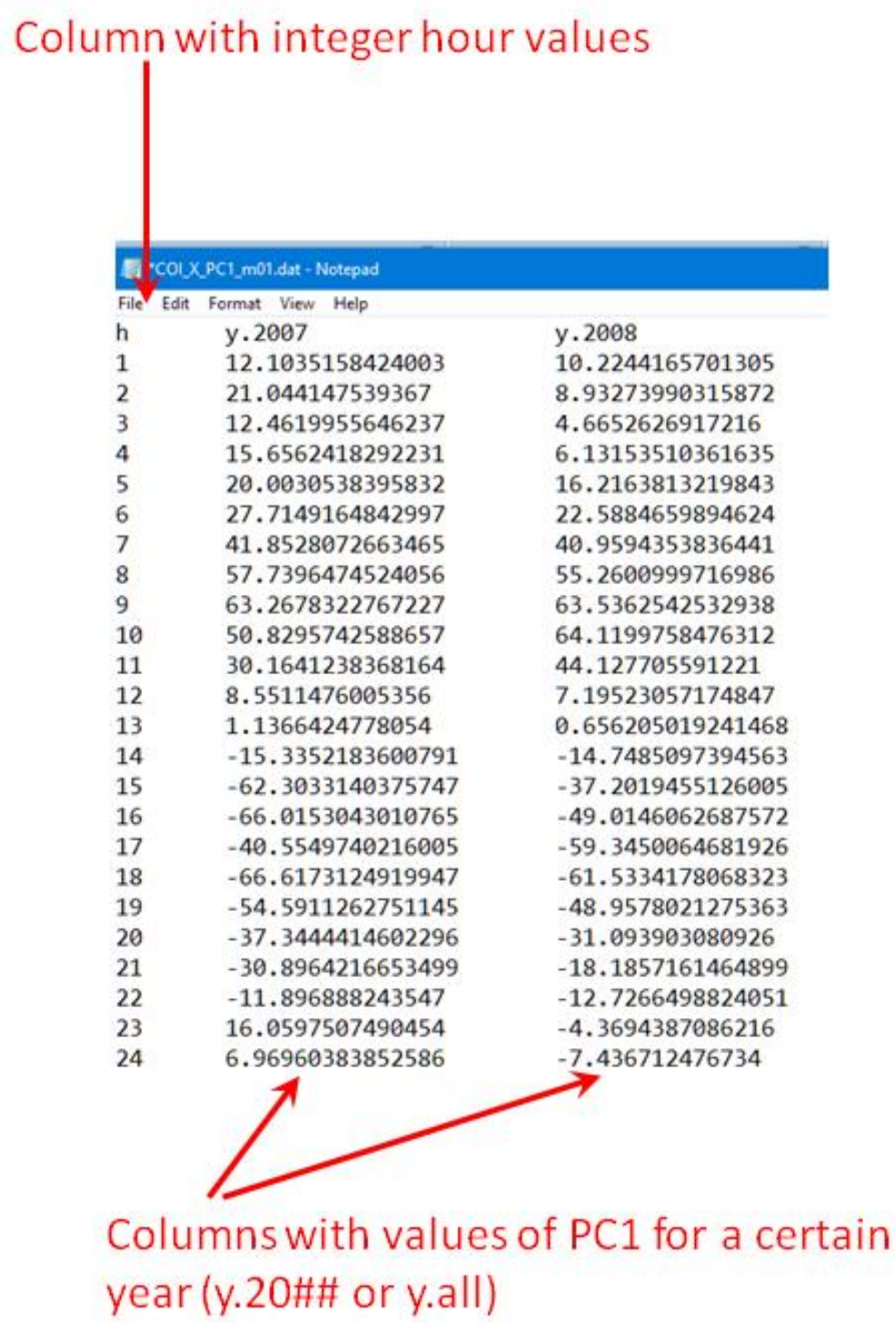

| h | y.2007 | y.2008 |
|---|---|---|
| 1 | 12.1035158424003 | 10.2244165701305 |
| 2 | 21.044147539367 | 8.93273990315872 |
| 3 | 12.4619955646237 | 4.6652626917216 |
| 4 | 15.6562418292231 | 6.13153510361635 |
| 5 | 20.0030538395832 | 16.2163813219843 |
| 6 | 27.7149164842997 | 22.5884659894624 |
| 7 | 41.8528072663465 | 40.9594353836441 |
| 8 | 57.7396474524056 | 55.2600999716986 |
| 9 | 63.2678322767227 | 63.5362542532938 |
| 10 | 50.8295742588657 | 64.1199758476312 |
| 11 | 30.1641238368164 | 44.127705591221 |
| 12 | 8.5511476005356 | 7.19523057174847 |
| 13 | 1.1366424778054 | 0.656205019241468 |
| 14 | -15.3352183600791 | -14.7485097394563 |
| 15 | -62.3033140375747 | -37.2019455126005 |
| 16 | -66.0153043010765 | -49.0146062687572 |
| 17 | -40.5549740216005 | -59.3450064681926 |
| 18 | -66.6173124919947 | -61.5334178068323 |
| 19 | -54.5911262751145 | -48.9578021275363 |
| 20 | -37.3444414602296 | -31.093903080926 |
| 21 | -30.8964216653499 | -18.1857161464899 |
| 22 | -11.896888243547 | -12.7266498824051 |
| 23 | 16.0597507490454 | -4.3694387086216 |
| 24 | 6.96960383852586 | -7.436712476734 |

**Figure 6.** Internal structure of the "COI_*_PC#_m##.dat" files.

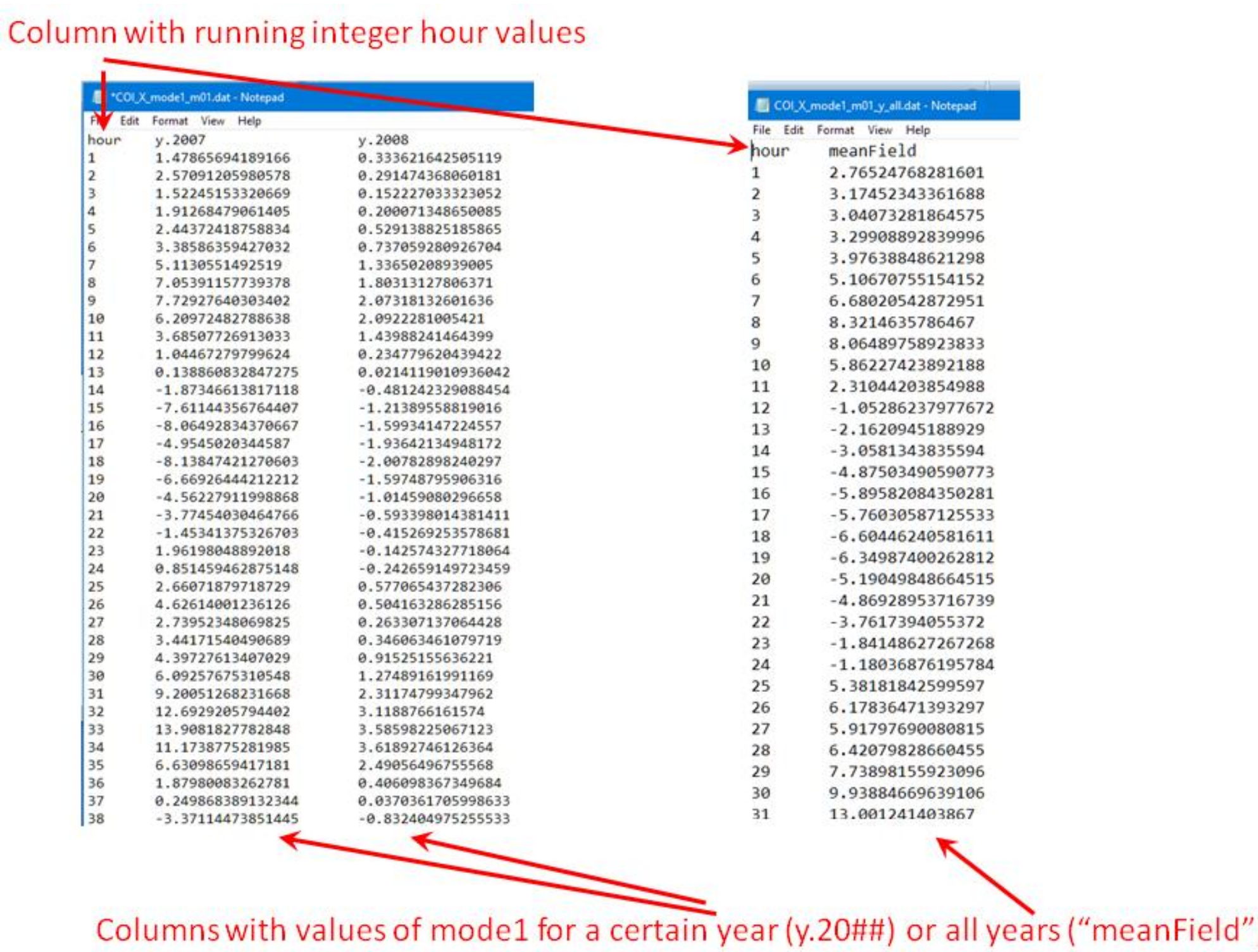

| hour | y.2007 | y.2008 |
|---|---|---|
| 1 | 1.47865694189166 | 0.333621642505119 |
| 2 | 2.57091205980578 | 0.291474368060181 |
| 3 | 1.52245153320669 | 0.152227033323052 |
| 4 | 1.91268479061405 | 0.200071348650085 |
| 5 | 2.44372418758834 | 0.529138825185865 |
| 6 | 3.38586359427032 | 0.737059280926704 |
| 7 | 5.1130551492519 | 1.33650208939005 |
| 8 | 7.05391157739378 | 1.80313127806371 |
| 9 | 7.72927640303402 | 2.07318132601636 |
| 10 | 6.20972482788638 | 2.0922281005421 |
| 11 | 3.68507726913033 | 1.43988241464399 |
| 12 | 1.04467279799624 | 0.234779620439422 |
| 13 | 0.138860832847275 | 0.0214119010936042 |
| 14 | -1.87346613817118 | -0.481242329088454 |
| 15 | -7.61144356764407 | -1.21389558819016 |
| 16 | -8.06492834370667 | -1.59934147224557 |
| 17 | -4.9545020344587 | -1.93642134948172 |
| 18 | -8.13847421270603 | -2.00782898240297 |
| 19 | -6.66926444212212 | -1.59748795906316 |
| 20 | -4.56227911998868 | -1.01459080296658 |
| 21 | -3.77454030464766 | -0.593398014381411 |
| 22 | -1.45341375326703 | -0.415269253578681 |
| 23 | 1.96198048892018 | -0.142574327718064 |
| 24 | 0.851459462875148 | -0.242659149723459 |
| 25 | 2.66071879718729 | 0.577065437282306 |
| 26 | 4.62614001236126 | 0.504163286285156 |
| 27 | 2.73952348069825 | 0.263307137064428 |
| 28 | 3.44171540490689 | 0.346063461079719 |
| 29 | 4.39727613407029 | 0.91525155636221 |
| 30 | 6.09257675310548 | 1.27489161991169 |
| 31 | 9.20051268231668 | 2.31174799347962 |
| 32 | 12.6929205794402 | 3.1188766161574 |
| 33 | 13.9081827782848 | 3.58598225067123 |
| 34 | 11.1738775281985 | 3.61892746126364 |
| 35 | 6.63098659417181 | 2.49056496755568 |
| 36 | 1.87980083262781 | 0.406098367349684 |
| 37 | 0.249868389132344 | 0.0370361705998633 |
| 38 | -3.37114473851445 | -0.832404975255533 |

| hour | meanField |
|---|---|
| 1 | 2.76524768281601 |
| 2 | 3.17452343361688 |
| 3 | 3.04073281864575 |
| 4 | 3.29908892839996 |
| 5 | 3.97638848621298 |
| 6 | 5.10670755154152 |
| 7 | 6.68020542872951 |
| 8 | 8.3214635786467 |
| 9 | 8.06489758923833 |
| 10 | 5.86227423892188 |
| 11 | 2.31044203854988 |
| 12 | -1.05286237977672 |
| 13 | -2.1620945188929 |
| 14 | -3.0581343835594 |
| 15 | -4.87503490590773 |
| 16 | -5.89582084350281 |
| 17 | -5.76030587125533 |
| 18 | -6.60446240581611 |
| 19 | -6.34987400262812 |
| 20 | -5.19049848664515 |
| 21 | -4.86928953716739 |
| 22 | -3.7617394055372 |
| 23 | -1.84148627267268 |
| 24 | -1.18036876195784 |
| 25 | 5.38181842599597 |
| 26 | 6.17836471393297 |
| 27 | 5.91797690080815 |
| 28 | 6.42079828660455 |
| 29 | 7.73898155923096 |
| 30 | 9.93884669639106 |
| 31 | 13.001241403867 |

**Figure 7.** The internal structure of the "COI_*_mode#_m##.dat" (left) and "COI_*_mode#_y_all_m##.dat" (right) files.

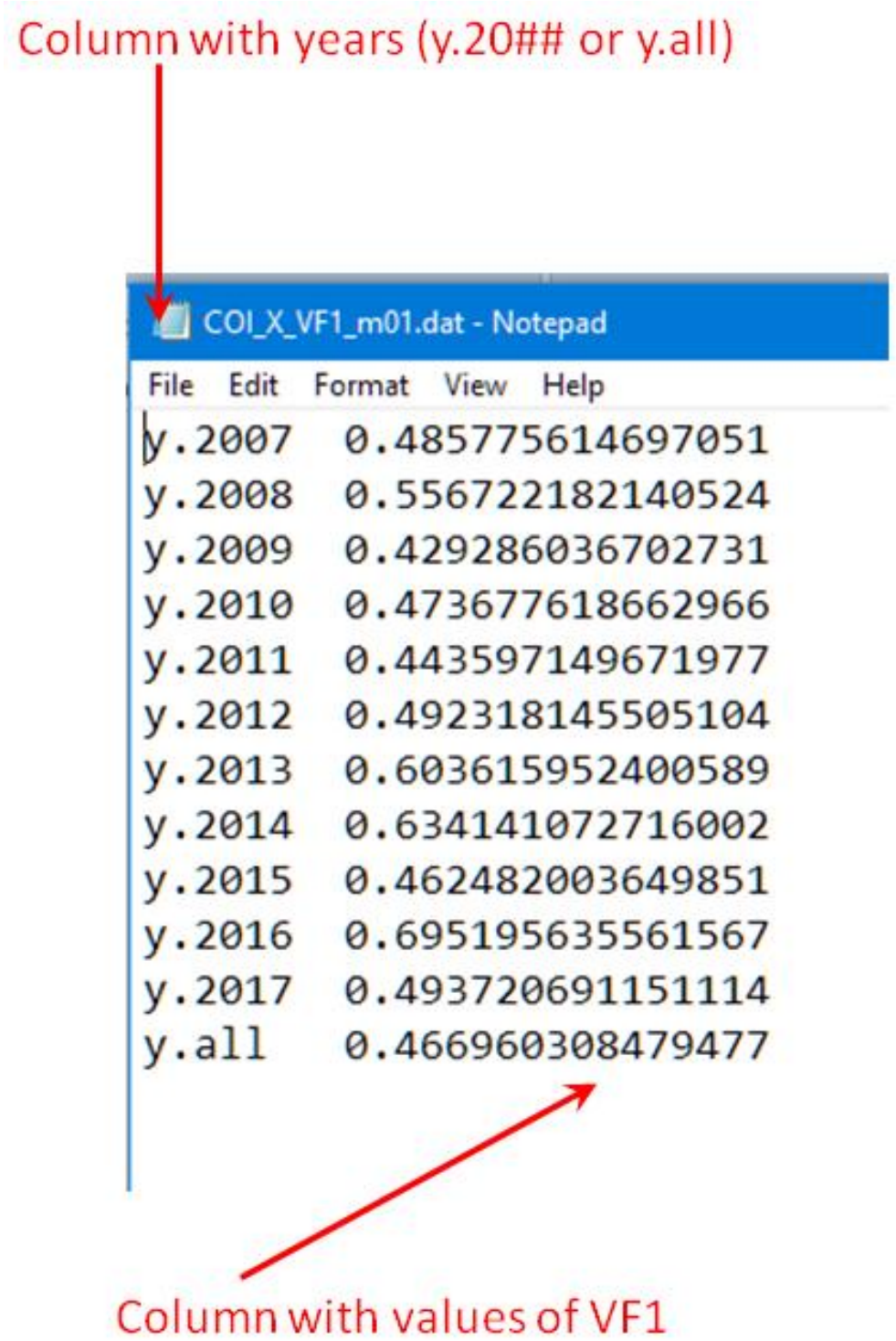

**Figure 8.** Internal structure of the “COI_*_VF#_m##.dat” files.

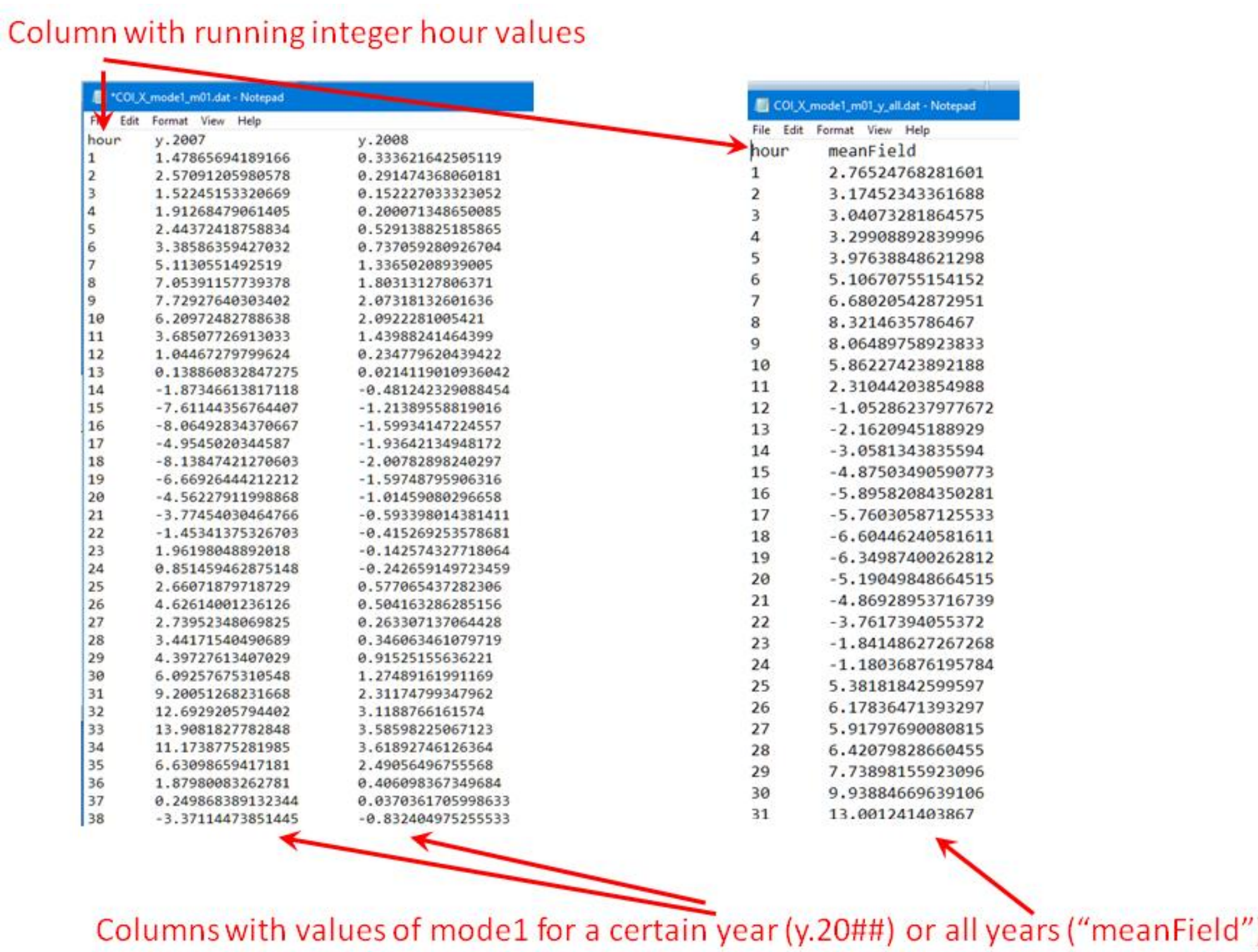

**Figure 9.** Internal structure of the “COI_*_meanField_m##.dat” (left) and “COI_*_meanField _m##_y_all.dat” (right) files.

The “plots\” folders contain two sub-folders named “mode 1” and “mode 2” that contain figures with corresponding plots of the EOFs (files named as “COI_2007_2017_*_LinInterp_m##_4PCA.EOF#.20##.png” for individual years and “COI_2007_2017_*_LinInterp_m##_4PCAEOF#11years.png” for for the all 11 years together) and PCs variations (files named as “COI_2007_2017_*_LinInterp_m##_4PCA.PC#.20##.png” for individual years and “COI_2007_2017_*_LinInterp_m##_4PCAPC#11years.png” for for the all 11 years together) for corresponding modes, where “*” is for “X”, “Y” or “Z”, “m##” is for a certain month and “20##” is for a certain year. Also, “plots\” folders contain plots showing all VF and the cumulative VF obtained for a specific month and a particular year: “COI_2007_2017_*_LinInterp_m##_4PCA.SCF#.20##.png” for individual years and “COI_2007_2017_*_LinInterp_m##_4PCASCF#11years.png” for all analyzed years – see Fig. 10.

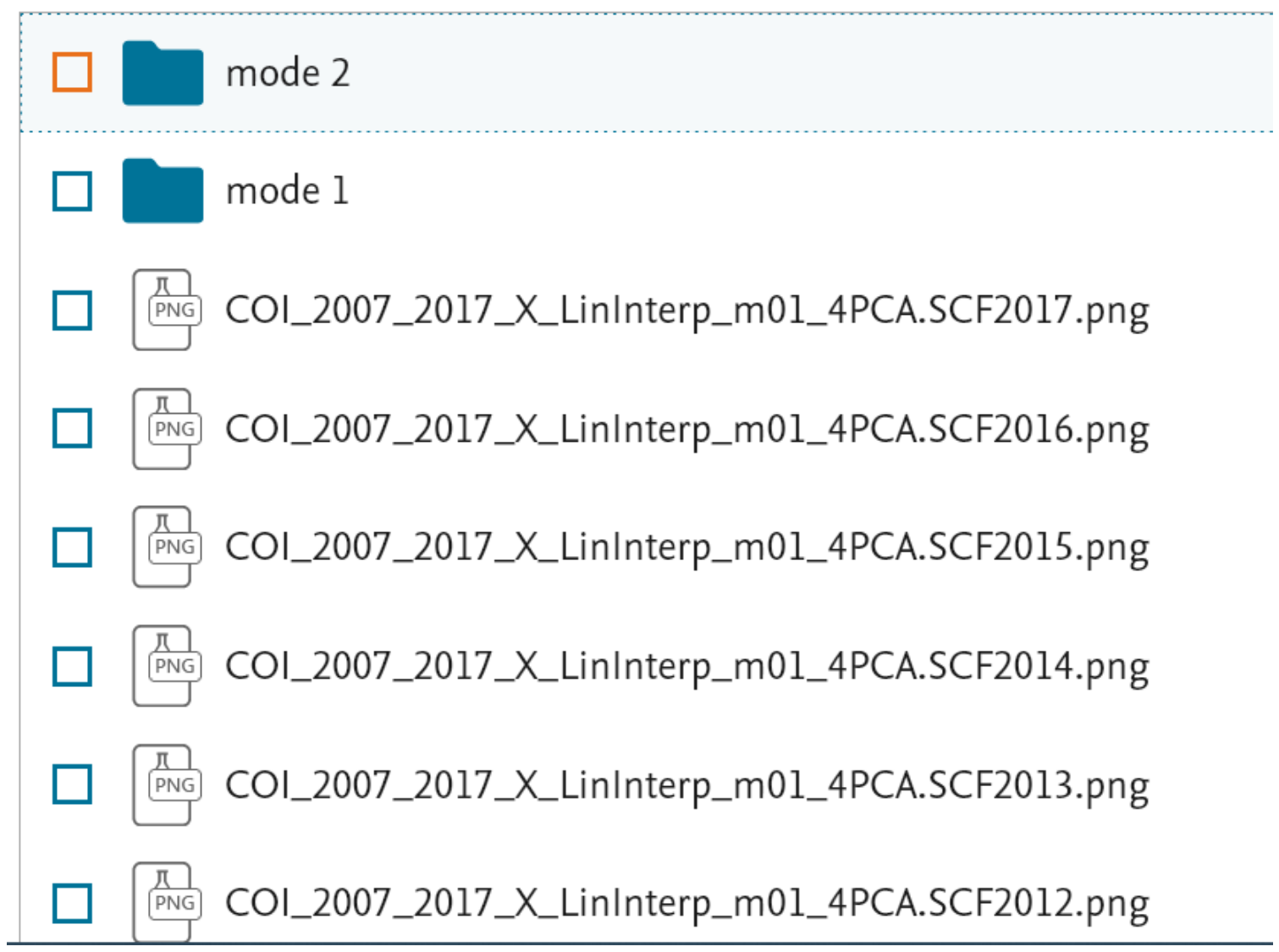

**Figure 10.** Examples of the data’s organization in monthly plots folders: “PCA modes\X\m01 – January\plots”.

For all of time series in this dataset, the time variable is either hours (PCs, S-type series, reconstructed modes) or days (EOFs, daily mean Field series). In all data files, the time variables are integer values. Conventionally, the hourly series of the geomagnetic parameters are centred not to the beginning of an hour but to its middle. Therefore, for the time plot of the series with one-hour time resolution, the correspondence between the integer hour and hour in time format (UTC, HH:MM) shown in Table 2 can be used.

**Experimental Design, Materials and Methods**

***Geomagnetic field components and their measurements***

Geomagnetic measurements at Coimbra Magnetic Observatory in Portugal (IAGA code COI) started in 1866 [1, 2]. In 2006 a new set of the absolute instruments was installed, providing good quality measurements of geomagnetic field components with 1-hour cadence [2]. Since there were no changes in the instruments or station location from 2006 to the present, the dataset obtained during this time interval can be considered homogeneous [2]. The detailed description of the COI instruments and metadata for the geomagnetic field components' series can be found in [1, 2]. The COI 1h geomagnetic data are regularly submitted to the World Data Centre for Geomagnetism and are available at its Geomagnetism Data Portal [3]. This dataset was used as raw data to obtain datasets for the regular geomagnetic field variations and main modes of the geomagnetic field presented in this paper.

**Table 2.** Correspondence between the integer hour and hour in the time format (UTC).

| Integer hour | Hour in UTC | Integer hour | Hour in UTC | Integer hour | Hour in UTC |
|---|---|---|---|---|---|
| 1 | 00:30 | 9 | 08:30 | 17 | 16:30 |
| 2 | 01:30 | 10 | 09:30 | 18 | 17:30 |
| 3 | 02:30 | 11 | 10:30 | 19 | 18:30 |
| 4 | 03:30 | 12 | 11:30 | 20 | 19:30 |
| 5 | 04:30 | 13 | 12:30 | 21 | 20:30 |
| 6 | 05:30 | 14 | 13:30 | 22 | 21:30 |
| 7 | 06:30 | 15 | 14:30 | 23 | 22:30 |
| 8 | 07:30 | 16 | 15:30 | 24 | 23:30 |

The geomagnetic field vector can be measured using a combination of three magnetic elements or components. These components are the total field (F) measured along the geomagnetic field direction at a particular point, horizontal component (H) measured along the magnetic meridian (positive in the direction of the N magnetic pole), declination (D), which is the angle between the magnetic and geographic meridians (positive eastward of true North), inclination (I) which is the angle between the horizontal plane and the F vector (positive downward), vertical component (Z, positive downward), and the north (X) and east (Y) components positive in the direction of the true (geographic) north and East, respectively.

For the relative instruments (i.e. variographs or variometers), the most widely used combinations are HDZ (cylindrical components) and XYZ (Cartesian). For the absolute, the combinations HDZ, HDI and DIF (spherical) are the most often used. Currently, DIF (absolute) and HDZ (relative) combinations are used at COI [2].

### *Methods to obtain regular variations of the geomagnetic field*

The daily or S-variations of the geomagnetic field are divided into two main classes: the "daily quiet" variation or Sq and the "daily disturbed" variation or SD (the name comes from the similarity of the form between the typical Dst-type and SD-type variations [4]). The datasets described in this paper present the series of S-type variations, Sq and SD, obtained from the 1h raw geomagnetic series for the X, Y and Z components of the geomagnetic field measured at COI during 11 years, from January 1, 2007, to December 31, 2017. The data are in nT. The time is in UTC, but the local time LT = GMT = UTC due to the COI location. The COI data have several gaps that were linearly interpolated in case of PCA. The S-type variations were extracted from the raw data using two different approaches described below for each of the X, Y and Z components separately.

***Quiet days Sq and SD***

The standard approach to calculate the Sq and SD variations is to use the so-called "geomagnetically quiet days" to select days of a month with the lowest geomagnetic activity level. In most cases, these "quiet days" are defined using the geomagnetic K-indices [5]. When local (obtained at a certain magnetic observatory) K-indices are used for the classification of a day, the resulting "quiet days" are the "local quiet days". When the planetary K-index (Kp) is used for the classification, the resulting "quiet days" are the "international quiet days" or IQD. In this work, we used IQDs routinely provided by the GFZ German Research Centre for Geosciences at the Helmholtz Centre in Potsdam, Germany [5].

Using the standard procedure, the Sq variation for a specific month is defined as the mean of daily variations of the month's five quietest days. In turn, the SD variation is calculated as a difference between the mean daily variations obtained using all days of a month (or S variation) and the corresponding Sq. Before the averaging, a baseline was removed from the raw daily series. There are two main ways to define the daily baseline for the geomagnetic field variations: the daily mean level and the night level (since under normal conditions, i.e. with no disturbances, the night is the time period with the lowest influence of the ionospheric currents on the ground measured geomagnetic field values). In this work, we used a baseline defined as a mean calculated for the night hours using the measurements made at 00:30 UTC, 01:30 UTC, 02:03 UTC, 03:30 UTC and 23:30 UTC of each day. Thus, the Sq variation values for the night hours are close to zero, and there are no significant differences between the night values of Sq at the beginning and the end of a day.

Each month was treated separately to take care of the seasonal variability of Sq variation. Thus, for each month of a year, we have one series of Sq and one series of SD variations, each consisting of 24 hourly values, overall 12*11 = 132 series for the Sq and SD variations, respectively. Also, we calculated the average Sq and SD variations for each month using all years of observation: e.g., using all January months from 2007 to 2017, etc., which additionally gives 12 series for the Sq and SD variations, respectively.

***Principal component analysis***

Principal component analysis (PCA) is a method allowing to extract main modes of variability of a series without any *a priory* assumption about the character of those variations (contrary to the widely used Fourier and wavelet analyses). An input data set is used to construct a covariance matrix and calculate corresponding eigenvalues and eigenvectors. The eigenvectors (empirical orthogonal functions, EOF) are used to calculate the principal components (PC). The combination of a PC and the corresponding EOF is called a "mode". Variations related to a certain mode can be reconstructed as a multiplication of a 1-column PC vector and a corresponding 1-row EOF vector. The eigenvalues allow estimation of the explained variances of the extracted modes. PCs are orthogonal and conventionally non-dimensional. The full descriptions of the method can be found in (e.g.) [6, 7, 8].

Recently, PCA was used to extract modes of the geomagnetic field's day-to-day variability, which were shown to be related to the S-type variations [9, 10, 11, 12, 13]. Here we applied a similar approach to extract modes of the geomagnetic field variations related to the regular variations on the daily time scale.

The PCA input matrices were constructed as follows: 24 rows for 24 hourly values per day and 28 to 31 columns (1 column for a day) depending on a month. All months were treated separately. All February matrices have a size 24 x 28. Individual input matrices (or data sets) were made for each of 12 months and each of 11 years (132 matrices). In addition, 12 matrices were constructed using the data for an individual month but with all years available (matrices with sizes 24 x 308, 24 x 330 or 24 x 341 depending on a month). Using this configuration of the input matrices, the principal components (PCs) correspond to daily variations of a different type that can be matched up with S-type variations calculated using the standard approach. The corresponding EOFs provide the amplitudes of a PC for each of the analyzed days. Also, PCA allows estimation of the "significance" of each of the extracted modes using their eigenvalues, a so-called variance fraction (VF) or squared covariance fraction (SCF) when the singular value decomposition method (SVD) is used to perform PCA, as in our cases. VF can be between 0 and 1 and when multiplied by 100% it shows the per cent of the total variability of the analyzed series related to a particular mode.

Only the three first PCs were selected to form the dataset presented in this paper. Overall, the first 3 PCA modes together explain from >60% to 98% of the COI X, Y and Z series variability depending on the month, year and the component.

## CRediT author statement

Anna Morozova: Conceptualization, Formal analysis, Methodology, Supervision, Validation, Software, Investigation, Visualization, Writing - original draft. Rania Rebbah: Software, Investigation, Data processing, Visualization, Writing - review & editing. Paulo Ribeiro: Data curation, Resources, Writing - review & editing

## Acknowledgments
CITEUC is funded by the National Funds through FCT (Foundation for Science and Technology) projects UID/00611/2020 and UIDP/00611/2020. This study is a contribution to the MAG-GIC project (PTDC/CTA-GEO/31744/2017).

## Declaration of Competing Interest
The authors declare that they have no known competing financial interests or personal relationships which have or could be perceived to have influenced the work reported in this article.